\documentclass[%
 reprint,
 amsmath,amssymb,
 aps,
nofootinbib,
]{revtex4-2}

\usepackage{graphicx}
\usepackage{amsmath}

\usepackage{color}

\usepackage{xcolor}

\usepackage{soul}

\begin{document}

\preprint{APS/123-QED}

\title{Compression and fracture of ordered and disordered droplet rafts}

\author{Pablo Eduardo Illing$^1$}
\email{pablo.eduardo.illing@emory.edu}

\author{Jean-Christophe Ono-dit-Biot$^2$}
\author{Kari Dalnoki-Veress$^{2,3}$}

\author{Eric R. Weeks$^1$}%
\email{erweeks@emory.edu}
\affiliation{$^1$ Department of Physics, Emory University, Atlanta, GA 30322, USA \\
$^2$ Department of Physics \& Astronomy, McMaster University, Hamilton, ON, L8S 4L8, Canada\\
$^3$ Gulliver, CNRS UMR 7083, ESPCI Paris, Univ. PSL, 75005 Paris, France.
}%


\date{\today}

\begin{abstract}
We simulate a two-dimensional array of droplets being compressed between two walls.  The droplets are adhesive due to an attractive depletion force.  As one wall moves toward the other, the droplet array is compressed and eventually induced to rearrange.  The rearrangement occurs via a fracture, where depletion bonds are quickly broken between a subset of droplets.  For monodisperse, hexagonally ordered droplet arrays, this fracture is preceded by a maximum force exerted on the walls, which drops rapidly after the fracture occurs.  In small droplet arrays a fracture is a single well-defined event, but for larger droplet arrays, competing fractures can be observed.  These are fractures nucleated nearly simultaneously in different locations.  Finally, we also study the compression of bidisperse droplet arrays.  The addition of a second droplet size further disrupts fracture events, showing differences between ideal crystalline arrays, crystalline arrays with a small number of defects, and fully amorphous arrays.  These results are in good agreement with previously published experiments.
\end{abstract}

\maketitle

\section{\label{sec:level1}Introduction}


Foams, emulsions, and colloids are often used as models for systems such as crystals and glasses \cite{ohern_Jamming_2003,weeks_Introduction_2017}.  Foams are gas bubbles in a liquid, with the gas-liquid interfaces stabilized by surfactant molecules.  Emulsions are similar in that they are droplets of one liquid in a second immiscible liquid, with surfactants stabilizing the liquid-liquid interfaces.  Colloids are composed of solid particles in a liquid.  The first published work using bubbles to model crystals was done by Bragg and Nye \cite{bragg_Dynamical_1947} and Bragg and Lomer \cite{bragg_Dynamical_1949}.  These soft matter systems can be used to study fundamental questions about order to disorder transitions \cite{zhang_CompressionInduced_2018,yunker_Observation_2010,goodrich_Solids_2014,keim_Role_2015,charbonneau_Glassy_2019,ono-dit-biot_Rearrangement_2020,berthier_Glass_2009,li_Assembly_2016,mari_Jamming_2009}, jamming \cite{bi_Jamming_2011,grob_Jamming_2014,illing_Mermin_2017,jorjadze_Attractive_2011,keim_Role_2015,mari_Jamming_2009}, and crystal nucleation and melting \cite{schaefer_Melting_1975,pusey_Phase_1986,wang_Imaging_2012,alsayed_Premelting_2005}.  More recently foams have also been used to study biological systems \cite{hayashi_surface_2004,gonzalez-rodriguez_soft_2012,pontani_biomimetic_2012,douezan_active_2011}. 

A key feature of these systems is their response to external stress. The mechanical response of these dispersions can be tuned by varying the composition \cite{irani_Impact_2014,grob_Jamming_2014,bonn_Yield_2017,golovkova_Depletion_2020,edwards_Foam_1990}.  It is also well known that materials become harder to deform as the volume fraction of the particulate phase is increased (that is, the colloidal particles, droplets, or bubbles, depending on the material) \cite{mason_Yielding_1996,mason_Elasticity_1995}.  Once a volume fraction threshold is reached, the system responds like a soft solid \cite{bonn_Yield_2017,mason_Yielding_1996,coussot_Coexistence_2002,nicolas_Deformation_2018}.  However, if enough force is applied, the system will plastically deform and flow. The relation between macroscopic flow and local plastic events has  been the focus of much work \cite{bonn_Yield_2017,durian_Foam_1995,durian_Scaling_1991,edwards_Foam_1990,friberg_Foams_2010}.

One concern using soft materials as models for crystals is that in contrast to atoms which are all identical, the components of a soft material are typically somewhat polydisperse.  Nonetheless, one can use low polydispersity (nominally ``monodisperse'') soft materials to model crystals \cite{bragg_Dynamical_1947,ono-dit-biot_Mechanical_2021}.  Such model systems allow one to investigate the effect of local plastic deformations and disorder \cite{ono-dit-biot_Rearrangement_2020,tong_Crystals_2015}, which  are connected to the bulk properties of the crystal, such as yield stress \cite{ono-dit-biot_Mechanical_2021,ono-dit-biot_Rearrangement_2020,yunker_Observation_2010,keim_Role_2015}.

The mechanical properties of a glass, like crystals, is heavily dependent on its microscopic structure, and this has been studied in a variety of soft materials serving as model glasses \cite{schall_Visualization_2004,vecchiolla_Dislocation_2019,alsayed_Premelting_2005,ohern_Jamming_2003,goodrich_Solids_2014,mari_Jamming_2009,babu_Excess_2016,zhang_CompressionInduced_2018,charbonneau_Glassy_2019,mizuno_Elastic_2013,ozawa_Role_2020,tong_Crystals_2015}.  Prior studies examined how the disordered structure of a glass affects a sample's mechanical properties \cite{yunker_Observation_2010,hanifpour_Mechanical_2014,yunker_Physics_2014,comtet_Atomic_2019}.  It is of interest to contrast crystals and glasses; for example numerical studies have shown that adding even a small amount of defects into a crystal drastically changes the mechanical properties of the resulting system \cite{ohern_Jamming_2003,goodrich_Solids_2014,mari_Jamming_2009,tong_Crystals_2015,babu_Excess_2016,zhang_CompressionInduced_2018,charbonneau_Glassy_2019,mizuno_Elastic_2013,ozawa_Role_2020}.

Experimental work by Ono-dit-Biot {\it et al.}~examined the ability of quasi-two-dimensional crystalline and noncrystalline samples to fracture under compression \cite{ono-dit-biot_Rearrangement_2020,ono-dit-biot_Mechanical_2021}.  The experiments consisted of a monolayer of oil droplets suspended in an aqueous solution.  The droplets packed into a raft held together by depletion forces.  The raft was then horizontally compressed between two parallel walls, causing the droplets to rearrange.  Nominally monodisperse rafts formed hexagonal close packed configurations.  During the compression process, the hexagonal packing would undergo coordinated fracture events.  Each such fracture allowed the crystal to reduce the number of rows by one, fitting into the narrower space imposed by the confining walls, while maintaining hexagonal order after the fracture event concluded.  However, when smaller droplets, which act as defects, were substituted into the droplet array, the coordinated fracture events were replaced by a series of smaller intermittent fractures.  With a sufficiently large number of defects, the samples behaved much more like a glass than like a disordered crystal.

The goal of this manuscript is to computationally replicate and extend the experimental results of Ono-dit-Biot {\it et al.} \cite{ono-dit-biot_Rearrangement_2020,ono-dit-biot_Mechanical_2021}. In particular, we expand on the prior results by simulating a larger number of droplets with a greater variety of starting configurations, allowing us to understand system size effects that were untested in the experiments.  Furthermore, we investigate the influence of experimental imperfections:  namely, the role of imperfectly parallel compressing walls, and understanding the role of polydispersity of particle sizes.  We also present analytic calculations which highlight the importance of attractive interactions between droplets to the observed results.

Our simulations include the three key forces present in the experimental work.  First, a repulsive force between the droplets (or the droplets and the walls) that is due to surface tension.  Second, an attractive force due to depletion from micelles  present in the experiments and modelled here with the Asakura-Oosawa model  \cite{asakura_Interaction_1954}.  Third, a dissipative viscous force acting on moving droplets, although this is minimal given the small velocities considered.  Our simulations reproduce the experimental observations, namely the fracture events and their dependence on the particle size distribution.  We also investigate a new phenomenon where the crystal fractures in multiple locations nearly simultaneously, disrupting the packing post-fracture, which occurs more frequently in larger droplet arrays.  Our work suggests that this phenomenon was likely suppressed in the experiments due to a slight tilt of the relative orientations of the two walls, taking them out of parallel by $\sim 0.2^\circ - 1.0^\circ$.

\section{Computational Methods}

\subsection{Simulation forces}

Our goal is to have a simulation which captures the key features of the prior work of Ono-dit-biot {\it et al.} \cite{ono-dit-biot_Rearrangement_2020,ono-dit-biot_Mechanical_2021}.  We use the Durian ``bubble model" \cite{durian_Foam_1995} to simulate the droplets' motions as the array is compressed.  In particular, we use the modified version presented by Tewari {\it et al.} in Ref.~\cite{tewari_Statistics_1999} which allows droplets to have different numbers of nearest neighbors; our code is specifically adapted from that used in Hong {\it et al.} \cite{hong_Clogging_2017,tao_Soft_2021}.
The athermal bubble model simulates the interactions between droplets in a viscous medium.  The model replaces the details of each droplet's deformation with a simple pairwise repulsive interaction.  The model assumes negligible inertial effects (appropriate for low-mass bubbles or slow-moving emulsion droplets), and viscous interactions.  For our work we add the effect of attractive depletion interactions between adjacent particles as well as between particles and the walls.

\begin{figure}[b]
\includegraphics[scale=0.55, origin=c]{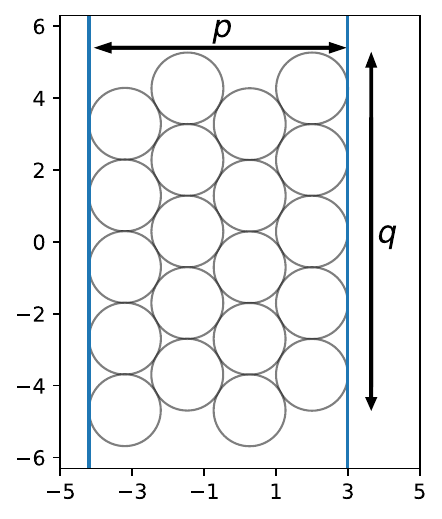}
\caption{\small Snapshots for a twenty droplet simulation. In this simulation the droplet is arranged in a $p=4$ by $q=5$ configuration.  } 
\label{fig:pxq}
\end{figure}

The first step in our simulation is to generate the droplets.  For nominally monodisperse simulations, we assign droplet radii according to a normal distribution with mean $\langle R \rangle=1$ and width $\delta = 1.25 \times 10^{-3}$.  This value is small enough to represent a single droplet size, while still allowing for some randomness that is inherent in the experiment.  For a bidisperse distribution, we generate droplets with radii $R_{\rm large}=1$ and $R_{\mathrm{small}}=0.765$, the same radius ratio as in the experimental work \cite{ono-dit-biot_Rearrangement_2020}. In all cases, the droplets are initially arranged in a hexagonal closed packed lattice, with $p$ rows and $q$ columns.  Rows are defined as a set of $q$ droplets aligned parallel to the walls.  An example is shown in Fig.~\ref{fig:pxq} with $p=4$ rows by $q=5$ columns.

Each droplet is modeled as a sphere, and the simulation starts by calculating all forces acting on each droplet.  The first is an elastic repulsive force between droplets.  If droplets $i$ and $j$ overlap, the repulsive force is:
\begin{equation}
    \vec{F}^{\mathrm{contact}}_{ij}=f_{\mathrm{0}}\left[ \frac{1}{|\vec{r}_i-\vec{r}_j|}-\frac{1}{|R_i+R_j|} \right]\vec{r}_{ij},
    \label{eq:bub-bub_contact}
\end{equation}
\noindent
where $R_i$ is the droplet radius, their positions are $\vec{r}_i$, and the difference vector is $\vec{r}_{ij}=\vec{r}_{i}-\vec{r}_{j}$.  An overlap occurs when a neighbor $j$ is close enough to the droplet $i$ such that $|\vec{r}_{ij}|<R_i+R_j$. Here, $f_0$ acts as a spring constant, the origin of which is the surface tension induced Laplace pressure.  In particular, Eqn.~\ref{eq:bub-bub_contact} avoids the need to simulate the actual deformation of the droplets by replacing the deformation with this effective force which is valid at low deformations, the regime of interest.   This linear (Hookean) response was observed in the experiments \cite{ono-dit-biot_Mechanical_2021}. 


The second interaction force between overlapping droplets is a viscous force, if the two droplets are moving at different velocities:
\begin{equation}
    \vec{F}_{ij}^{\mathrm{viscous}}=b(\vec{v}_i-\vec{v}_j)
    \label{eq:viscous_bubint}
\end{equation}
\noindent
with $b$ being the viscous coefficient, and $\vec{v}_i$ the velocity of a given particle.  This force acts on each droplet in a direction that tries to bring their velocities into agreement: for example, if droplet $i$ is motionless then the viscous force from droplet $j$ acting on $i$ is in the direction of $v_j$.

A final important force in the experiment is the depletion force:  an attractive force acting between droplets that are sufficiently close.  In the experiment, this is due to small surfactant micelles.  In the simulation this is modeled as an effective force between neighboring droplets, which in this case are droplets with distance $r_{ij}<R_i+R_j+2a_s$. Here $a_s$ is the radius of the depletant which thus sets the range of the attractive interaction.  The depletion forces are calculated using the Asakura-Oosawa model \cite{asakura_Interaction_1954}.  The first step needed to calculate the depletion interactions is to calculate the overlapping volume between a pair of spheres  with a radius $R'_i\equiv R_i+a_s$ \cite{crocker_Entropic_1999}:
\begin{equation}
    \begin{split}
        &V_{\mathrm{overlap}}(r_{ij},R_i',R_j')=\frac{\pi}{12r_{ij}}\left(R'_i+R'_j-r_{ij}\right)^2 \times\\
        & \left(r_{ij}^2 +2r_{ij}\left(R'_i+R'_j\right)-3({R'_{i}}^2+{R'_{j}}^2)+6R'_i R'_j\right).
    \end{split}
    \label{eq:Voverlap}
\end{equation}

Using the overlap volume we can obtain the associated Helmholtz free energy, which we then differentiate to get the depletion force:
\begin{equation}
    \vec{F}_{ij}^{\mathrm{dep}}=\frac{\phi_c}{8a^3_{s}}\frac{\partial V_{\mathrm{overlap}}(r_{ij},R'_i,R'_j)}{\partial r_{ij}} \hat{r}_{ij},
    \label{eq:depletion_bub}
\end{equation}
\noindent
where the direction of the force is attractive between the two particles.  In this formula, $\phi_c$ is a constant related to the temperature and volume fraction of the depletant.  In the simulation we set $a_s=\frac{\langle R \rangle}{20}=1/20$.  The formula for $\partial V_{\mathrm{overlap}}/\partial r_{ij}$ is:
\begin{equation}
    \begin{split}
        \frac{\partial V_{\mathrm{overlap}}(r_{ij},R'_i,R'_j)}{\partial r_{ij}}= \\ 
        \frac{\pi}{4} \left[ r_{ij}^2 - 2({R'_i}^2+{R'_j}^2) + 
        \left(\frac{{R'_i}^2 - {R'_j}^2}{r_{ij}}\right)^2 \right].
    \end{split}
    \label{dVoverlap}
\end{equation}

In addition to calculating the droplet-droplet interactions, we also need to calculate the droplet-wall interactions for droplets sufficiently close to the wall. The repulsive force from the wall is given by
\begin{equation}
    \vec{F}^{\mathrm{wall,repel}}_i=f_0\langle R \rangle \left(r_{\mathrm{wall},i}^{-1}-R_i^{-1}\right) \hat{n}_{\mathrm{wall}},
    \label{eq:rep_wall}
\end{equation}
where $r_{{\mathrm{wall},i}}$ is the distance from the droplet center to the wall.  Note this force diverges as $r_{\mathrm{wall}} \rightarrow 0$, preventing any droplets from passing through the wall.  In particular this form differs from Eqn.~\ref{eq:bub-bub_contact} by using the unit normal vector $\hat{n}_{\mathrm{wall}}$ rather than $\vec{r}_{\mathrm{wall}}$, which is what leads to the divergence.  The magnitude of the attractive depletion force between a droplet and the wall is given by:
\begin{equation}
F^{\mathrm{wall, dep}}_{i}= \frac{\pi \phi_c}{8 a_s^3} (R_i+2 a_s -r_{\mathrm{wall},i}) (R_i + r_{\mathrm{wall}})
    \label{eq:dep_wall}
\end{equation}
for every droplet with $r_{\mathrm{wall},i}<R_i+2a_s$.
We will summarize these two terms into:
\begin{equation*}
    F^{\mathrm{wall}}_i=F^{\mathrm{wall,repel}}_{i}+F^{\mathrm{wall, dep}}_i,
\end{equation*}
noting that the two components point in opposite directions (and thus $F^{\mathrm{wall}}_i$ can be zero if these two components are in balance).

The Durian Bubble Model is originally for massless bubbles \cite{durian_Foam_1995}, and in our situation we treat droplets in a low Reynolds number limit for which inertial effects are negligible.  Accordingly, the net force is always zero; the velocity of each particle is always such that the velocity-dependent viscous forces balance the other forces.  Thus, we combine Eqns.~\ref{eq:bub-bub_contact}, \ref{eq:viscous_bubint}, and \ref{eq:depletion_bub} and solve for the velocity:
\begin{equation}
    \vec{v}_i=\langle \vec{v}_j \rangle+ \frac{1}{bN_i}\sum_j\left( F_{ij}^{\mathrm{contact}}-F_{ij}^{\mathrm{dep}}\right) \hat{r}_{ij}+\frac{1}{b}F_{\mathrm{wall}} \hat{n}_{\mathrm{wall}},
    \label{eq:Durian_bub}
\end{equation}
where $N_i$ are the total number of neighbors for particle $i$.  We use fourth order Runge-Kutta to solve this differential equation for the velocities at each time step.

\subsection{Model parameters}\label{SS:mp}

The model sketched above has many parameters.  In this section we discuss how these parameters are set based on comparison with the prior experimental work (Refs.~\cite{ono-dit-biot_Mechanical_2021,ono-dit-biot_Rearrangement_2020}) and on computational convenience.  We start by fixing $b=1$, $\langle R \rangle = 1$, $v_{\mathrm{wall}}= 10^{-4}$, $\phi_c=10^{-4}$, $f_0=10$, and $a_s= 0.05$.  Several nondimensional ratios allow comparisons to the experiment.

First, the range of the depletion force is given by the ratio of the size of the micelles and the size of the droplets. With mean droplet radius $\langle R\rangle \approx 20$~$\mathrm{\mu m}$, and depletion micelles which have size $a_s\approx 5$~$\mathrm{nm}$, the experimental range is $\alpha_1^{\mathrm{expt}} \approx 2.5 \times 10^{-4}$. In the simulations the range of the depletion interaction is set to:
\begin{equation}
    \alpha_1^{\mathrm{sim}} = \frac{a_s}{\langle R \rangle}=0.05.
\end{equation}
Here $\alpha_1^{\mathrm{sim}}$ is larger than $\alpha_1^{\mathrm{expt}}$, although still much less than 1. This choice avoids numerical instabilities which would occur if the depletion force was too short range.

Second, we need to understand the relative importance of the depletion and viscous forces.  In the experiment, this ratio of forces is
\begin{equation*}
    \alpha_2^{\mathrm{expt}} = \frac{v_{\mathrm{wall}}\eta}{W}\approx 10^{-5},
\end{equation*}
\noindent where the viscosity $\eta\approx 1\times 10^{-3}$~Pa$\cdot$s, the depletion energy per unit area between two droplets $W\approx 1\times 10^{-6}\mathrm{J/m^2}$, and the speed of the wall $v_{\mathrm{wall}}\approx 3\times 10^{-7}\mathrm{m/s}$ \cite{ono-dit-biot_Mechanical_2021}.  In the simulation, the same ratio is
\begin{equation}
    \alpha_2^{\mathrm{sim}} = \frac{b v_{\mathrm{wall}}a_s}{\phi_c} = 0.05.
    \label{eq:alfa2}
\end{equation}

\noindent In both the simulation and in the original experiments, the depletion force is stronger.  That being said, in the simulations, the effect of viscosity is more significant than in the experiment.  This choice is to keep the simulation computational time reasonable; reducing the viscosity coefficient $b$ would require a smaller integration time step.

Finally we compare the forces of repulsion and depletion in the experiment:
\begin{equation*}
    \alpha_3^{\mathrm{expt}} = \frac{k}{W}\sim 10^4,
\end{equation*}
where $k\approx 10^{-3} \mathrm{N/m} $ is the spring constant associated with the oil droplets' surface tension \cite{mason_Elasticity_1995,rehfeld_Adsorption_1967}. In our simulations we have:
\begin{equation*}
    \alpha_3^{\mathrm{sim}} = \frac{f_0 a_s}{\phi_c} = 0.5 \times 10^{4}.
\end{equation*}
These are the same order of magnitude; the factor of $0.5$ difference means that the simulated droplets are slightly softer than the experimental droplets.  Adjusting the ratio in the simulation would again increase the computational costs, so we judge our parameter choices to be a reasonable compromise between computational costs and adequately capturing the experimental limits (short range attractive forces, small viscous forces compared to depletion, large repulsive forces compared to depletion).

\subsection{Simulation timescales}

We need to choose the simulation time step carefully to allow for the correct integration of all interactions.  As shown in Eqns.~\ref{eq:bub-bub_contact}, \ref{eq:viscous_bubint}, and \ref{eq:depletion_bub}, the magnitude of the different forces are set by the constants $f_{\mathrm{0}}, \phi_c$, and $b$ for repulsive, depletion, and viscous forces respectively. From these constants, together with the speed of the walls, $v_{\mathrm{wall}}$, average droplet radii, $\langle R \rangle$ and the depletant radius, $a_s$, we can define three different time scales:  $\tau_1=\frac{\langle R \rangle b}{f_0} = 10^{-1}$, $\tau_2=b a_s^2/\phi_c = 25$, and $\tau_3=\langle R \rangle/v_{\mathrm{wall}} = 10^4$.  $\tau_1$ is the time scale for two overlapping droplets to push apart in the absence of the depletion force, and is the fastest time scale.  $\tau_2$ is the time scale for two overlapping droplets to pull together due to the depletion force, which is slower than $\tau_1$ because as noted above, the depletion force is weaker than the repulsive force.  Finally, $\tau_3$ is the time scale for the walls to move a distance $\langle R \rangle$.  Given these results, we set the simulation time step to be $\Delta t = \tau_1 = 0.1$.  We have checked that simulations run with smaller time steps give the same results as those run with $\Delta t = 0.1$.  The implication of $\tau_1 \ll \tau_3$ is that during rearrangements, the walls will move a negligible distance.

\subsection{Simulation goals}

Given that the simulation parameters are chosen to match the experiment to a reasonable extent, it is worth stating what we wish to replicate.  The experiment measures forces exerted on the moving walls, and relate these forces to the effective spring constant of two interacting droplets due to the depletion force \cite{ono-dit-biot_Mechanical_2021}.  Our wall forces likewise must relate to the effective spring constant in our simulation, so we consider this in Secs.~\ref{analytic} and \ref{results}.  This is not a replication {\it per se} so much as allowing us to illuminate the importance of both compressive and tensile forces acting between droplets.  That being said, one important experimental observation to replicate is the relationship between the wall forces, array size, and effective spring constant acting between a pair of droplets, which will be examined in Sec.~\ref{results}-A.

One experimental observation to replicate is how the forces on the walls are changed when the droplet array has a mixture of particle sizes.  When the experimental droplet array consisted of a nearly $50/50$ mixture of large and small droplets, the array rearranged in a nearly continuous sequence of small fracture events; however, this observation was limited to a 23-droplet array \cite{ono-dit-biot_Rearrangement_2020}.  We wish to replicate the observations and extend them to larger array sizes.

Finally, moving beyond replication, we will examine how the fractures depend on the system size, droplet polydispersity, and wall tilt angles:  factors that are easier to vary smoothly in the simulation as compared to experiment.

\section{Analytical results}
\label{analytic}

We wish to understand the force required to compress the droplet array.  We start by considering the effective spring constant between two droplets.  We then consider compressing three droplets.  Due to the attractive depletion force, compressing three droplets requires one effective spring to be stretched while the other two are compressed.  In this section we take $R_i = R_j = R = 1$.

\subsection{Effective spring constant:  two droplets}

For two droplets in contact the balance of repulsive and attractive forces in equilibrium lead to a harmonic interaction with an effective spring constant. Balancing Eqns.~\ref{eq:bub-bub_contact} and \ref{eq:depletion_bub}, the equilibrium distance between two particles 
can be approximated as:
\begin{equation}
    d_{\rm eq} = 2R - 2 a_s\left(\frac{L^2}{2R^2} - 1\right)^{-1}.
    \label{eq:req}
\end{equation}
using:
\begin{equation}
    L^2=\frac{8f_0a^3}{\pi \phi_c}
\end{equation}
We can use two of the nondimensional ratios analyzed in Sec.~\ref{SS:mp},
the range of the depletion forces $\alpha_1$ and the ratio of repulsion to depletion forces $\alpha_3$, to write   
\begin{equation}
    \frac{L^2}{R^2} =  \frac{8}{\pi}\alpha_1^2\alpha_3 = \frac{100}{\pi} = 31.8.
    \label{eq:L2}
\end{equation}
By substitution of Eqn.~\ref{eq:L2} into Eqn.~\ref{eq:req}, we see that the term multiplying $a_s$ has a small value for our simulations so we will define:
\begin{equation}
    \epsilon = \left(\frac{L^2}{2R^2} - 1\right)^{-1}\approx 0.0671.
\end{equation}
We can then finally write the equilibrium position as:
\begin{equation}
    d_{\rm eq} = 2(R - a_s \epsilon)\approx 2(1-0.00336)\approx 1.993.
    \label{eq:reqfinal}
\end{equation}
As expected the equilibrium position would be at $2R$ if depletion wasn't present.  With depletion, the equilibrium position is adjusted by a small fraction of the depletant radius $a_s$.  With the parameters used in the simulations the particles overlap but just slightly.  (In the experiment, this implies that the droplets would be slightly deformed due to the depletion force.  Given that $\alpha_3^{\rm expt} \sim 10^4$, the experimental deformation is likely unobservable.)

At this point we can calculate the effective spring constant response for monodisperse droplet-droplet interactions due the balance of depletion and repulsion. Using both Eqn.~\ref{eq:bub-bub_contact} and \ref{eq:depletion_bub}, and doing a small displacement from equilibrium $\Delta r_{ij}$, results in the force increasing by:

\begin{equation}
    \Delta F= \left(\frac{f_0 }{2R}-\frac{R \pi \phi_c}{8 a_s^3}\right)\Delta r_{ij}=k_1 \Delta r_{ij} ,
    \end{equation}
which leads to
\begin{equation}
    k_1 \approx 4.69
    \label{eq:k1}
\end{equation}
\noindent
as the effective spring constant for droplet-droplet interactions.  The depletion force slightly reduces the spring constant from that due purely to repulsion, which is $f_0/(2R) = 5$.

We can similarly calculate the energy associated with breaking a depletion bond. In this case we must calculate the work needed to separate two droplets from their equilibrium separation, $d_{\mathrm{eq}}$, up to the the point where depletion turns off, $d_{\mathrm{off}}=R_i+R_j+2a_s$.
\begin{equation*}
    E_{\mathrm{depletion}}=-\frac{  \phi_c}{8 a_s^3}\int_{d_{\mathrm{eq}}}^{d_{\mathrm{off}}}\frac{\partial V_{\mathrm{overlap}}}{\partial r_{ij}} dr_{ij}=\frac{ \phi_c}{8 a_s^3}\left.V_{\mathrm{overlap}}\right\vert^{d_{\mathrm{eq}}}_{d_{\mathrm{off}}}
\end{equation*}

\noindent The minus sign is due to the fact that the motion to separate the droplets opposes the depletion force. Since there is no overlapping volume at $d_{\mathrm{off}}=R_i+R_j+2a_s$, we have only the equilibrium volume:
\begin{equation}
    E_{\mathrm{depletion}}=\frac{ \phi_c}{8 a_s^3}V_{\mathrm{overlap}}(d_{\mathrm{eq}}).
    \label{eq:E1pre}
\end{equation}

\noindent However since the equilibrium distance is less than the radius of the droplets, we must also take into account the repulsive force's work, which assists in separating the droplets:
\begin{equation}
    W_{\mathrm{rep}}=f_0 \int_{d_{\mathrm{eq}}}^{R_i+R_j}\left( 1- \frac{r_{ij}}{R_i+R_j}\right) dr_{ij},
    \label{eq:W1}
\end{equation}

\noindent and so the corrected term for the energy needed to break a bond between two droplets is the difference between Eqns.~\ref{eq:E1pre} and \ref{eq:W1}:
\begin{equation}
        E_1=\frac{ \phi_c}{8 a_s^3}V_{\mathrm{overlap}}(d_{\mathrm{eq}})-\frac{f_0 (R_i+R_j-d_{\mathrm{eq}})^2}{2(R_i+R_j)}.
\end{equation}

\noindent Using Eqns.~\ref{eq:req} and \ref{eq:L2}, we can simplify this further to:
\begin{equation}
        E_1=\frac{\phi_c}{8 a_s^3}V_{\mathrm{overlap}}(d_{\mathrm{eq}})-\frac{2f_0 a_s^2 \epsilon^2}{R}.
        \label{eq:E1}
\end{equation}

\noindent We can now replace all the values by the corresponding constants, and for $d_{\mathrm{eq}}$ and $L^2$ from Eqns.~\ref{eq:req} and \ref{eq:L2}, respectively, which gives us the energy stored per bond in the monodisperse case:
\begin{equation}
E_1=0.00173.
\label{eq:E1value}
\end{equation}

We can repeat a similar calculation for the effective spring constants and bond energy present at the walls, and obtain:
\begin{equation}
    k^{\mathrm{wall}}_{\mathrm{eff}}=9.57 = 2.04 k_1, 
    \label{eq:kwall}
\end{equation}
\noindent and
\begin{equation}
    E^{\mathrm{wall}}_{1}=0.00340 = 1.965 E_1.
\end{equation}


\subsection{Equivalent spring model for three droplets}
\label{three}

\begin{figure}[b]
\includegraphics[scale=0.5, origin=c]{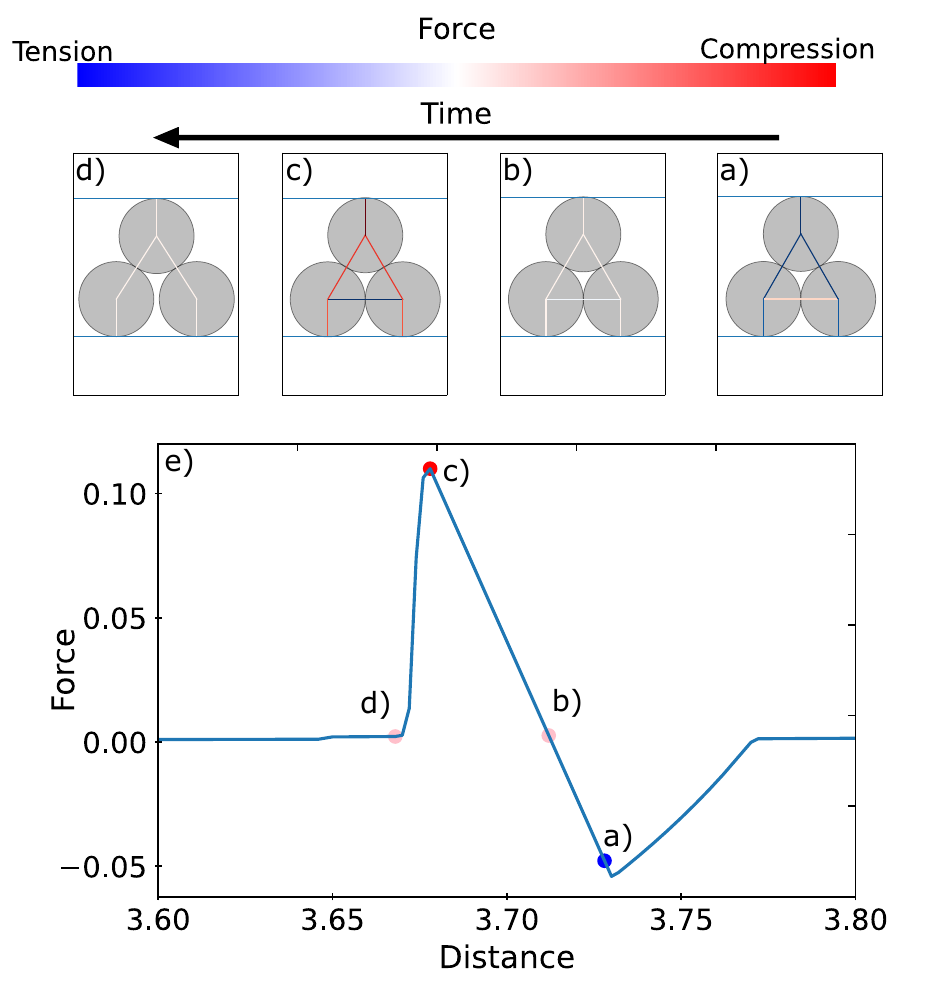}
\caption{\small  (a-d):  Consecutive snapshots of the row reduction for three droplets:  note that time increases from right to left, to match panel (e).  Blue bonds are under tension, whereas red indicates compression. In panel (a), as the walls move together, the three droplets attach to the walls due to depletion and thus exert tension on the walls, resulting in the negative force peak seen in panel (e) at distance $(2+\sqrt{3})R \approx 3.73$. As the top wall continues to move closer to the bottom wall, the droplets go through equilibrium [panel (b)], and eventually the walls begin compressing the droplets [panel (c)], leading to rise in the force on the walls.  At the peak of the force, the horizontal bond between the two left particles is under tension.  When this bond between the two bottom droplets breaks, the force rapidly drops [panel (d)].}
\label{fig:3bub}
\end{figure}




We next consider the compression of three monodisperse droplets in an equilateral triangle arrangement, shown schematically in Fig.~\ref{fig:3bub}(a). In this case we start with two rows of droplets and the compression causes a rearrangement to one row. As the top wall moves towards the bottom wall, the droplet cluster attaches to the two walls due to depletion forces.  Initially the bonds to the walls are all under tension due to the depletion force, pulling on the two walls. As the distance between the walls continues to decrease, the droplets go through equilibrium (panel b) with the spacing between each pair of droplets being $d_{\mathrm{eq}}$ (Eqn.~\ref{eq:req}).  As the compression proceeds, the force continues to rise as the distance between the walls decreases (panel c).  During this process, the diagonal bonds compress while the horizontal bond between the two droplets on the bottom wall is under tension.  Eventually this horizontal bond breaks, which allows the diagonal compressed bonds to relax (panel d); from this point onward the droplets will continue to move with only the viscosity resisting their motion until they are reduced to a single row of droplets (not shown). In Fig.~\ref{fig:3bub}(e) we plot the force on the walls as a function of the distance. Time is advancing from right to left. The linear rise of the force from position (a) to (c) indicates that the the array is compressed elastically, until the vertical bond breaks at (c) and the row reduction occurs.

\begin{figure}
\includegraphics[scale=0.6, origin=c]{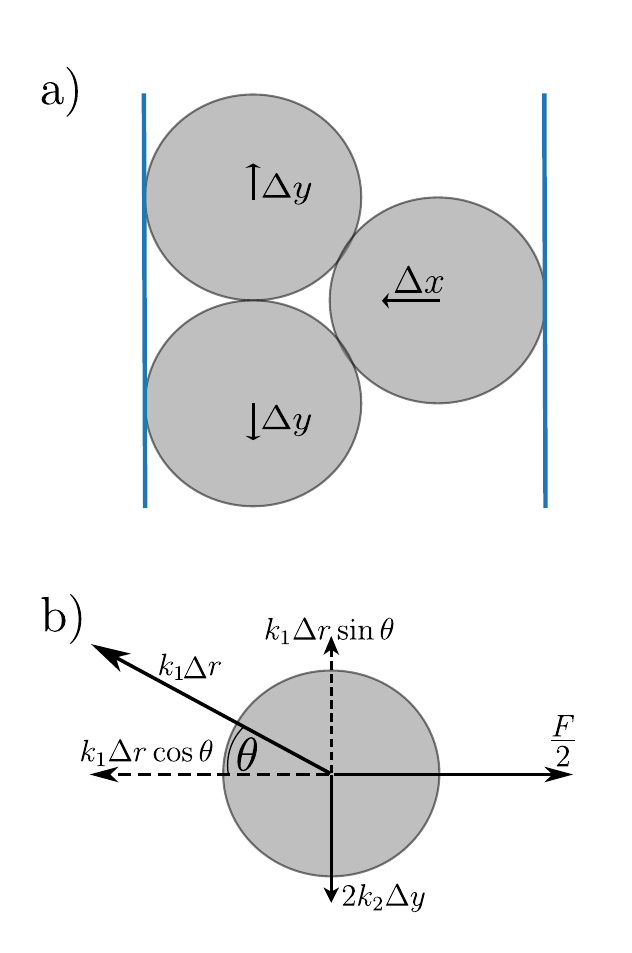}
\caption{\small {(a) Sketch of small displacements of three droplets.  (b) Free body diagram of forces acting on the top left droplet.  $F/2$ is from the wall, with the other $F/2$ contribution acting on the bottom droplet.  The vertical spring is stretched by $2 \Delta y$, so accordingly the force indicated as $2 k_2 \Delta y$ is a tension force from the bottom droplet.  The $k_1 \Delta r$ force is a compression force from the right side droplet.  As the droplets are monodisperse $\theta = 30^\circ$.  Changes in $\theta$ for small $\Delta x$ and $\Delta y$ can be ignored to first order.}}
\label{fig:freebody}
\end{figure}

To explain the elastic rise in force [position a) to c) in Fig.~\ref{fig:3bub}(e)], we note that each droplet bond has spring constant $k_1$, which can be used to calculate the equivalent spring constant for the triangular array.  For these three droplets, the compression force between the two left droplets and the left wall is half that of the  compression force of the single droplet on the right and the adjacent wall.   To find a relationship between the force $F$ exerted by the walls on the droplet pack and the horizontal displacement of the walls from the equilibrium position (when $F=0$), we start by considering the situation sketched in Fig.~\ref{fig:freebody}(a): the left two droplets only move in the $y$ direction, with the top droplet moving up by $\Delta y$ and the bottom moving down by $\Delta y$; and the right droplet moves left by $\Delta x$ under the action of the force $F$.  (To be clear, this is in the reference frame where the left droplets do not move horizontally.  In practice, all three droplets move horizontally under the influence of the walls, with the net horizontal displacement between the right droplet and the left droplets as $\Delta x$.)


A free body diagram for the top droplet is shown in Fig.~\ref{fig:freebody}(b).  For the moment we will consider the vertical bond between the two left droplets to have spring constant $k_2$ in order to illustrate the role of tension, but since the spring constant is the same (for small displacements) whether under tension or compression, we will eventually set $k_2 = k_1$.  The distance between the top droplet and right droplet is initially $d_{\mathrm{eq}}$.  When the droplets begin to move, the change in this distance is 
\begin{equation}
    \Delta r \approx \Delta x \cos \theta - \Delta y \sin \theta .
    \label{deltar}
\end{equation}
This expression is valid in the limit where $\Delta x,\Delta y \ll r_{eq}$, and changes to the angle $\theta$ due to droplet movement can be ignored as they are a second order correction.  This formula for $\Delta r$ has been chosen with signs so that $\Delta r > 0$ when the droplets are being compressed, consistent with the direction of the force indicated in Fig.~\ref{fig:freebody}(b).  Balancing the two forces in the vertical direction gives
\begin{equation}
    \Delta y = \frac{k_1 \sin \theta \cos \theta}{k_1 \sin^2\theta + 2 k_2} \Delta x,
    \label{eq:deltay}
\end{equation}
which relates the horizontal and vertical displacements.

If we take the extreme case where $k_2 \rightarrow \infty$, Eqn.~\ref{eq:deltay} gives us $\Delta y \rightarrow 0$, as no matter how much we push, the two vertical droplets are stuck together at a fixed separation.  On the other hand if there is no adhesion force, then $k_2=0$ and $\Delta y=\cot \theta \Delta x$, which corresponds to the droplets displacing as needed to accommodate the right droplet moving leftwards (and thus keeping $\Delta r = 0$).
Finally for the case we consider in the simulations, $k_1=k_2$ and $\theta = 30^\circ$, giving:
\begin{equation}
    \Delta y=\frac{\sqrt{3}}{9} \Delta x \equiv C \Delta x.
    \label{eq:deltay2}
\end{equation}
This is indeed the relation between the displacements that we observe in the three droplet simulation. 

Substitution of Eqn.~\ref{eq:deltay2} into Eqn.~\ref{deltar} for $\Delta r$, and balancing the horizontal forces in Fig.~\ref{fig:freebody}(b), leads  to
\begin{equation}
    F = 2 k_1 \cos \theta (\cos \theta - C \sin \theta) \Delta x.
\end{equation}
This expression relates the wall force $F$ to the compression $\Delta x$ of the three droplets; the term in front of $\Delta x$ is an effective spring constant equal to $(4/3) k_1$.  Note that the term with $C$, which allows for the vertical motion of the two left side droplets, reduces the effective spring constant slightly (as the ratio of the second term to the first is $C \tan \theta = 1/9$).  Again to understand the role of the tension force between the two vertically oriented droplets, we consider the limits for $C$.  If $k_2 \rightarrow \infty$ then $C=0$ and the effective spring constant is equal to $(3/2)k_1$.  Intriguingly, this is the result one gets for two springs in parallel, in series with one spring, which apart from the diagonal connections is what we see in Fig.~\ref{fig:freebody}(a).  In contrast, if $k_2 = 0$, then $C = \cot \theta$ and the term in parentheses would be zero, thus resulting in $F = 0$.  Thus the tension bond plays an important role in generating the wall force.

In addition to the spring-like interaction of the three compressed droplets, there are also spring-like interactions with the walls.  At the left side wall, because there are two droplets, the effective wall interaction behaves with spring constant of $2 k_{\rm wall}$; on the right side we have simply $k_{\rm wall}$.  These three springs act in series, so thus the overall effective spring constant the system has is
\begin{equation}
    \frac{1}{k_{\rm eq}} = \frac{1}{2 k_{\rm wall}} + \frac{1}{k_{\rm wall}} + \frac{1}{2 k_1(\cos^2 \theta - C \sin \theta \cos \theta)}
    \label{eq:k3}
\end{equation}
which simplifies to $0.673 k_1 = 3.16$ using Eqns.~\ref{eq:k1} and \ref{eq:kwall}.  This is exactly the value we measure from the slope of the elastic regime in Fig.~\ref{fig:3bub}(e), matching to the three significant figures we have been using.  The close agreement is perhaps a bit surprising, given that the analytic calculation has been assuming small displacements whereas the simulation uses the full form of the depletion interaction.  The agreement also confirms that our wall speed is slow enough that viscous forces are not adding significantly to our measured wall force.

We can also relate the energy required to break one depletion bond (Eqn.~\ref{eq:E1}) to the force peak $F_p$.  Ignoring energy stored in compressive interactions, the elastic energy $k_{\mathrm{eq}}\Delta x^2/2$ gets converted into breaking $\Delta n$ bonds, so we have
\begin{equation}
    E_1 \Delta n = \frac{1}{2} k_{\mathrm{eq}} \Delta x^2 = \frac{F_{\rm max}^2}{2 k_{\rm eq}}.
    \label{eq:e1}
\end{equation}
From the data in Fig.~\ref{fig:3bub}, we have $\Delta n = 1, k_{\rm eq}=3.16$, and $F_{\rm max} = 0.110$ leading to $E_1= 0.00191$ which is 10\% larger than the prediction of Eqn.~\ref{eq:E1}.  The discrepancy is precisely because of the extra energy stored in the compressed droplets.  The extra energy gets converted into droplet motion once the bonds under tension are broken.  

\section{Computational results for large droplet arrays}
\label{results}

\subsection{Equivalent spring model for nominally monodisperse crystals}
\label{monodisperse}

\begin{figure}
\includegraphics[scale=0.45, origin=c]{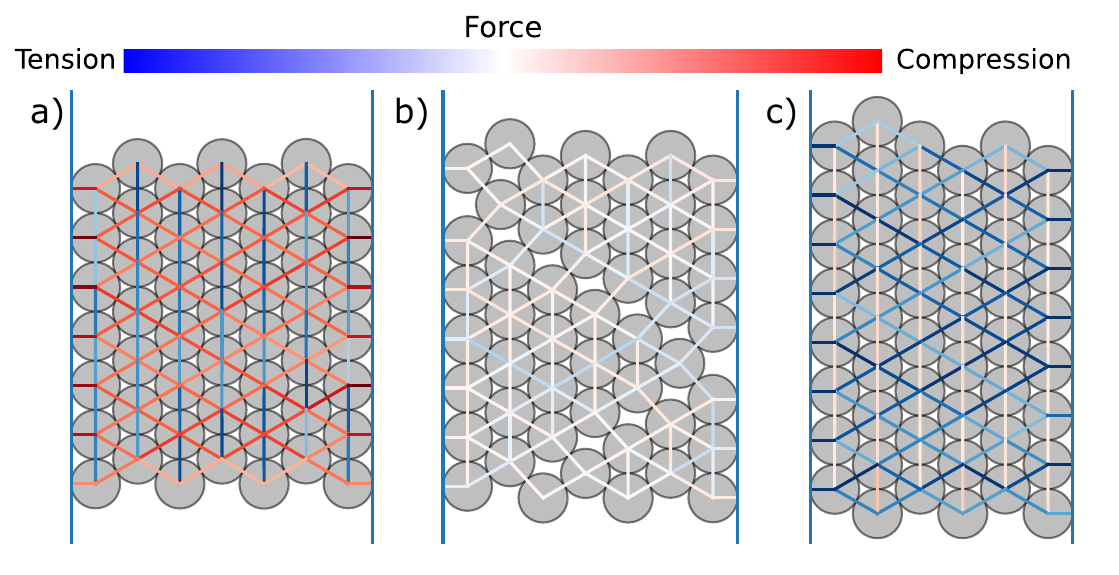}
\caption{\small {A $7\times 7$ droplet array in its initial configuration undergoing a row reduction. In (a) we can see that the array is being compressed between the walls, as evidenced by the red compression bonds. The blue tension bonds run parallel to the walls, with the depletion force preventing the crystal from spreading along the direction of the walls. In (b) a global fracture has occurred, splitting the crystal into four separate pieces and relaxing the forces; many droplets are close to their equilibrium separation distance. In (c) the crystal settles into a $6\times 8$ hexagonal-close-packed configuration and the depletion forces pull the droplets and walls closer together. The extra droplet is in the second column. Movie S1 in the Supplemental Material depicts the compression process for the array shown in this figure \cite{Supplemental}.}}
\label{fig:global_fracture}
\end{figure}

We will next consider the general case of a nominally monodisperse rectangular array of droplets with $p$ rows and $q$ columns.  We will consider the specific example of a $7\times 7$ array but also, where relevant, discuss results from simulations with other numbers of droplets.  Figure \ref{fig:global_fracture} shows this $7 \times 7$ droplet array undergoing a row reduction from $p=7$ to $p=6$ rows.  A video of the compression process for this particular simulation is available in the Supplemental Materials Movie S1 \cite{Supplemental}. All simulations are initialized by placing the droplets in a perfectly ordered array; one such initial state is shown in Fig.~\ref{fig:global_fracture}(a), which corresponds to the system before a fracture.  The red lines indicate compression forces and the blue lines indicate tension forces.  The tension is caused by the depletion forces between droplets along the vertical direction (parallel to the walls), therefore separating the droplets apart, pulling against the depletion force which holds the crystal together. 
In Fig.~\ref{fig:global_fracture}(b) the global fracture has broken the crystals into four distinct pieces. The forces decrease in magnitude, indicated by the light pink and light blue lines, showing the compression and tension in the crystal has been relieved during the fracture.  Each piece moves as an essentially solid assembly; the relative position of the droplets within this assembly does not matter, and droplets at the boundaries move similarly to those in the middle.  Finally in Fig.~\ref{fig:global_fracture}(c) a new crystal with $p=6$ rows forms, with tension pulling the droplets back into a hexagonal configuration.  In this case the walls experience a tension force from the attractive depletion bonds which are not yet at their equilibrium position.

\begin{figure}[b]
\includegraphics[scale=0.6, origin=c]{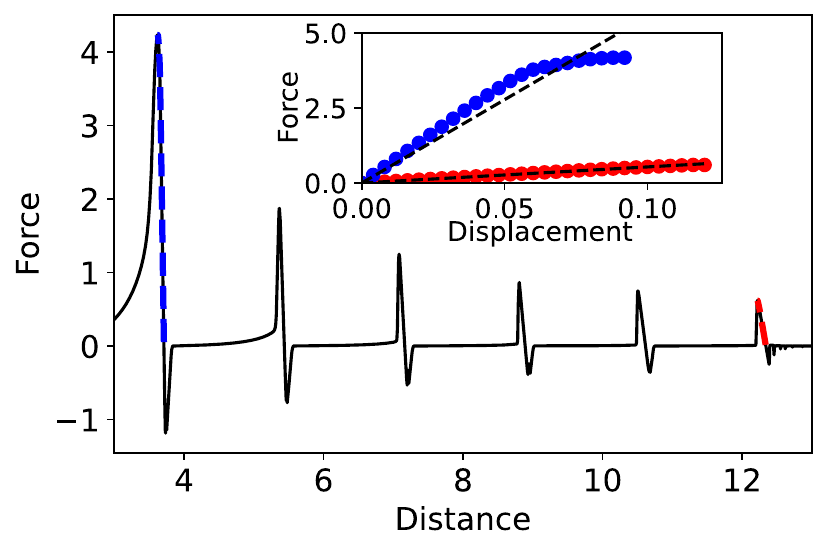}
\caption{\small {Force on the left wall as a function of the distance between the walls for a $7 \times 7$ monodisperse droplet array.  The distance starts large and decreases, so time increases from right to left. Each successive peak is taller than the last, due to the increased amount of bonds that need to be broken. 
The blue and red slope are used to obtain $k_{eq}$, as in seen in the inset. }}
\label{fig:forcelft_graph}
\end{figure}

Clearly during the compression process the force exerted by the droplet packing on the walls varies in both magnitude and direction.  In Fig.~\ref{fig:forcelft_graph} we plot the force exerted on the left wall as the crystal is compressed and undergoes row reduction.  Note that time increases from right to left, as the horizontal axis is the distance between the walls which decreases with time.  We wish to understand the features of this graph, and will start with the easiest:  during the compression force minima ($F<0$) occur. These correspond to the droplets nestling into a new hexagonal arrangement, being pulled in together by depletion forces, and pulling on the wall as the droplets settle into this more compact arrangement. An example of this  corresponds to Fig.~\ref{fig:global_fracture}(c) where all the bonds perpendicular to the wall are under tension:  the more compact configuration exerts this tension on the walls due to the attractive depletion force.

We next turn to the question discussed previously for two and three droplets:  what is the effective spring constant of this droplet array?  The inset in Fig.~\ref{fig:forcelft_graph} shows the increase of the force from zero as a function of the compression from the equilibrium position for the first row reduction (red) and the last (blue).  The dashed lines in the inset show the linear fit used to obtain the $k_{eq}$ for that row transition.  We can see that once the walls begin compressing the crystal the force rises monotonically, with the crystal responding elastically until finally a catastrophic fracture event occurs. This is due to the tension forces being sufficient to break the depletion bonds between the droplets along the fracture.  At large compression (blue data points) the force is less than expected, as when the array is down to two layers, being compressed into one layer, the bonds at the ends of the array break first and relieve some of the wall force while the bonds in the middle are still intact.  
That is, the fracture does not occur everywhere simultaneously. 

The effective spring constant is larger when the droplet array has fewer rows and more columns.  This can be understood by a generalization of the spring model to bigger arrays.  We have a rectangular array of $p$ rows (parallel to the walls) and $q$ columns; the equivalent spring constant is therefore that of a matrix of $p$ springs in series and $q$ springs in parallel.  The interactions with the two walls, with spring constant $k_{\rm wall} \approx 2 k_1$, has the effect of adding an additional row.  This leads to the following equation that relates the equivalent spring constant $k_{\mathrm{eq}}$ to the spring constant of a single droplet $k_1$:

\begin{equation}
    k_{\rm eq}=\frac{q}{p+1} k_1
    \label{eq:k1pq}
\end{equation}
\noindent In general Eqn.~\ref{eq:k1pq} is a simplification as it ignores the effect of the springs under tension, as described for three droplets in Sec.~\ref{three}.  Nonetheless this is a useful approximation.  For the red slope in Fig.~\ref{fig:forcelft_graph} we get $k_1=5.27$, and for the blue slope we get $k_1=7.65$.  The higher $k_1$ for the blue data is because at this point the droplet array is quite wide and to compress the array requires nontrivial motions at the edges of the array as will be discussed below.  These large edge motions lead to viscous forces which increase $k_{eq}$ and thus the measured $k_1$. These measured values for $k_1$ are about 12-15\% larger than the true value of $k_1$, illustrating the enhanced elasticity due to the tension bonds.

\begin{figure}
\begin{center}
\includegraphics[scale=0.6, origin=c]{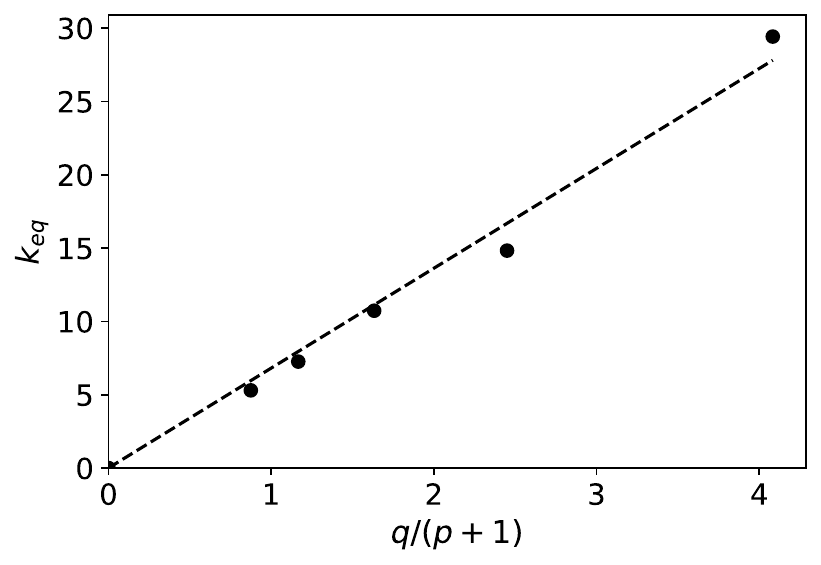}
\caption{\small {Linear fit for the compression of the equivalent spring constants, as a function of the aspect ratio of the array $q/(p+1)$ for 49 droplet array. Using Eqn.~\ref{eq:k1pq} and the linear fit gives us the spring constant for a single droplet, $k_1=7.08 \pm 0.6$.}}
\label{fig:compression_spring_fit}
\end{center}
\end{figure}

For each row reduction we perform a linear regression on the force as a function of compression distance and obtain the corresponding value of the equivalent spring $\Delta F=k_{eq} \Delta x$.  The graph of $k_{eq}$ is shown in Fig.~\ref{fig:compression_spring_fit} and is linear as a function of $q/(p+1)$ as predicted by Eq.~\ref{eq:k1pq}.  As the crystal is compressed $q/(p+1)$ grows and therefore $k_{eq}$ grows as we have a larger number of springs arranged in parallel and fewer in series. The linear fit of this graph gives $k_1 = 7.08 \pm 0.5$, which is higher than our theoretical $k_1 = 4.69$ (Eq.~\ref{eq:k1}).  This procedure was repeated for different runs with the same crystal configuration (thus different realizations of our slight polydispersity), as well as for arrays containing $20$ to $400$ particles, obtaining a mean value of $k_1\approx 6.1$. As with the individual measurement of $k_{eq}$ discussed above, the likely cause of the higher $k_1$ is due to the viscous forces acting on the larger droplet array, as well as the breakdown of the approximations used to calculate $k_1$.

To understand how the viscous forces affect our system, consider a row reduction transitioning from $p$ to $p-1$ rows, for example as shown in Fig.~\ref{fig:global_fracture} with $p=7$ initially.  The number of columns $q$ is a function of the total number of droplets $N$ and $p$, and thus increases from $q=N/p$ to $q'=N/(p-1)$. We will continue by analysing the displacement of one of the droplets at the edge of the configuration -- that is, at the top or bottom of a row. Before the row reduction the rows of the array have length $2R q$, and $2R q'$ afterwards. Taking the center of the array to be our origin, the displacement of an edge droplet during this row reduction is:
\begin{equation}
    d_{\mathrm{edge}}=\frac{N}{p-1}\frac{2R}{2}-\frac{N}{p}\frac{2R}{2}
    \label{eq:dedge}
\end{equation}
where we have divided by two as the array expands symmetrically around the origin.
The time needed for this transition to occur is the time needed for the walls to move the distance of one row, $t=\sqrt{3}R/v_{\mathrm{wall}}$. Dividing Eqn.~\ref{eq:dedge} by $t$ we obtain:
\begin{equation}
    v_{\mathrm{edge}}=\frac{v_{\mathrm{wall}}}{\sqrt{3}} \left( \frac{N}{p(p-1)} \right).   
    \label{eq:vedge}
\end{equation}
Based on Eqn.~\ref{eq:vedge} we can see that the speed of an edge droplet depends on the size of the array, as well as which transition it is. Replacing Eqn.~\ref{eq:vedge} into Eqn.~\ref{eq:alfa2} for $\alpha_2^{\mathrm{sim}}$ we have:
\begin{equation}
.  
    \alpha_2^{\mathrm{edge}} = \frac{1}{\sqrt{3}} \left( \frac{N}{p (p-1)} \right) \alpha_2^{\rm sim}.  
     \label{eq:alfa2edge}
\end{equation}
For the first transition in Fig.~\ref{fig:global_fracture}, we have $N=49$, $p=7$ and thus $\alpha_2^{\rm edge} = \alpha_2^{\rm sim}  49/(42\sqrt{3}) \approx 2 \alpha_2^{\rm sim}$.  For the last row reduction starting with $p=2$, this becomes $\alpha_2^{\rm edge} = \alpha_2^{\rm sim}  49/(2\sqrt{3}) \approx 14 \alpha_2^{\rm sim} \approx 0.7$.  This shows that during the last row reduction, for the edge droplets the viscous forces are now comparable to the depletion forces, even for an array of modest size with $N=49$.

We verified this computationally using $N=49$ and using half and double our usual value of $v_{\rm wall}$.  As expected, the simulations running at double the wall speed had more significant viscous effects for the last row reductions, while the simulations running at half the wall speed had less noticeable viscous effects.

The next feature of Fig.~\ref{fig:forcelft_graph} to explain is the peaks in the force.  As the number of rows is reduced and the number of columns increases, Fig.~\ref{fig:forcelft_graph} shows the force required for the fracturing increases significantly.  This is because more depletion bonds need to be broken.

As we did previously for the three droplet case, we can obtain the depletion energy per bond from the force peaks for each transition, continuing from Eqn.~\ref{eq:e1} which we can rewrite as:
\begin{equation}
    F_{\rm max}=\sqrt{2 E_1 k_{\mathrm{eq}} \Delta n}
    \label{eq:Fckeq}
\end{equation}
\noindent where $\Delta n$ is the number of bonds broken during the row reduction. To rewrite this equation we will use Eqn.~ \ref{eq:k1pq} to replace $k_{\mathrm{eq}}$ with $k_1$ and $N=p \times q$ to eliminate $q$.  We will additionally assume that the number of broken bonds per transition is $\Delta n=2 q$, which is true when the array fractures into equilateral triangles, as was the case in the original experiments.  This leads to:
\begin{equation}
    F_{\rm max}=2 N\sqrt{\frac{k_1 E_1}{p^3+p^2}}
    \label{eq:Fc_perfect}
\end{equation}

\begin{figure}
\includegraphics[scale=0.6, origin=c]{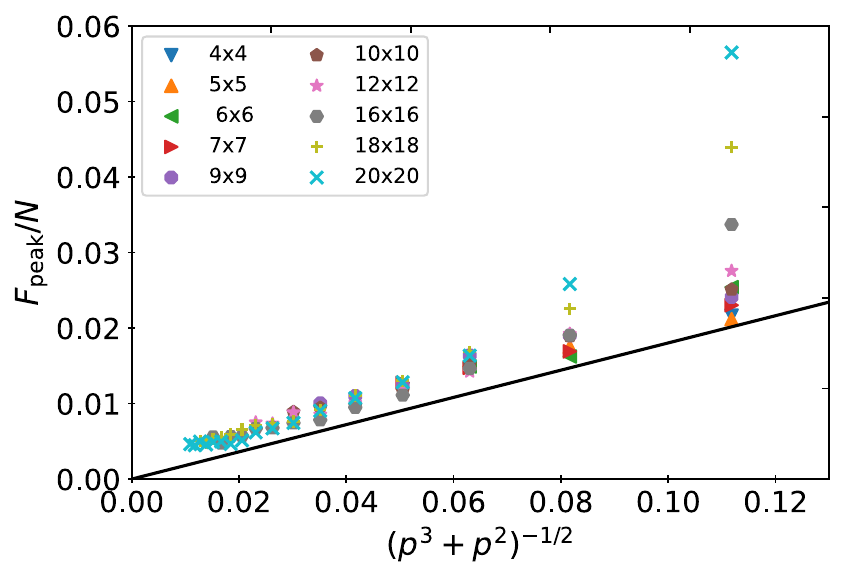}
\caption{\small {Evolution of the normalized peak height as a function of $(p^3+p^2)^{-1/2}$ for a different variety of starting configurations. The solid line represent the values predicted by Eqn.~\ref{eq:Fc_perfect}.   The discrepancies at the right side are due to the increasing influence of viscous forces, which become more significant for large arrays with small $p$.  The right-most data correspond to $p=4$.}}
\label{fig:normpeak_high}
\end{figure}

\noindent To test this we plot in Fig.~\ref{fig:normpeak_high} the force peaks $F_{\rm max}$ divided by the total number of droplets $N$ in each simulation as a function of the $(p^3+p^2)^{-1/2}$ for a wide range of $N$ and $p$.  The solid line shows the prediction of Eqn.~\ref{eq:Fc_perfect} using $k_1$ and $E_1$ from Eqs.~\ref{eq:k1} and \ref{eq:e1}.

Figure \ref{fig:normpeak_high} shows that most force peaks exceed the expected value corresponding to Eq.~\ref{eq:Fc_perfect}. There are several reasons for this difference.  First, Eq.~\ref{eq:Fc_perfect} uses $k_1$ which neglects the influence of the tension bonds.  Second, viscosity dissipates some of the energy the walls put into the system, which is more significant for bigger arrays and when the arrays have fewer rows $p$ [thus higher $(p^3+p^2)^{-1/2}$].  With fewer rows, the bubbles at the ends of the array must move faster to reach the new configuration, while the wall keeps moving at the same speed; see Eqn.~\ref{eq:alfa2edge}.  Equation \ref{eq:alfa2edge} also shows that larger arrays (larger $N$) have larger viscous forces, in agreement with what is seen in Fig.~\ref{fig:normpeak_high}.  Third, there are situations where $\Delta n > 2 q$ (caused by more complex fracture events) which will be discussed in Sec.~\ref{competing}, which thus increases $F_{max}$.

To summarize, we have successfully replicated the experimental observation that each successive row reduction requires greater compression, as there are more depletion bonds that need to be broken \cite{ono-dit-biot_Rearrangement_2020}.  Like the experiments, we successfully relate the spring constant of a single droplet to the array aspect ratio dependence of the wall forces \cite{ono-dit-biot_Mechanical_2021}.  Our results also illuminate the influence of viscosity (in Fig.~\ref{fig:normpeak_high}), which is more observable in the simulations due to the larger nondimensional number $\alpha_2^{\rm sim}$ (Eqn.~\ref{eq:alfa2}).

The equivalent spring model is therefore a useful tool for understanding the characteristics of a nominally monodisperse droplet array as it is compressed.  In the next section we will take a closer look at the behavior of arrays which are no longer considered monodisperse.

\subsection{Bidisperse Aggregates}

As seen in the previous section, a raft made up of low polydisperse droplets is a model crystalline packing.  In this section we introduce defects and increase the polydispersity of the simulated samples to study these new aggregates during compression, which more closely resemble glassy materials.

We start by analyzing the behaviour of bidisperse aggregates. In these arrays the particles can have a radius of either $R=1$ or $R=0.765$ (to match the experiments of Ref.~\cite{ono-dit-biot_Rearrangement_2020}). We define the defect fraction $\phi$ as:
\begin{equation}
    \phi=N_{\mathrm{small}}/N_{\mathrm{total}}
    \label{eq:defect_fraction}
\end{equation}
where $N_{\mathrm{small}}$ is the number of smaller droplets in the aggregate, and $N_{\mathrm{total}}$ is the total number of droplets. The defect fraction $\phi$ varies from zero to one. In the case where $\phi=0$ or $\phi=1$, we return to the monodisperse case seen in the previous section, corresponding to a crystal made exclusively of large or small droplets.

\begin{figure}
\includegraphics[scale=0.55, origin=c]{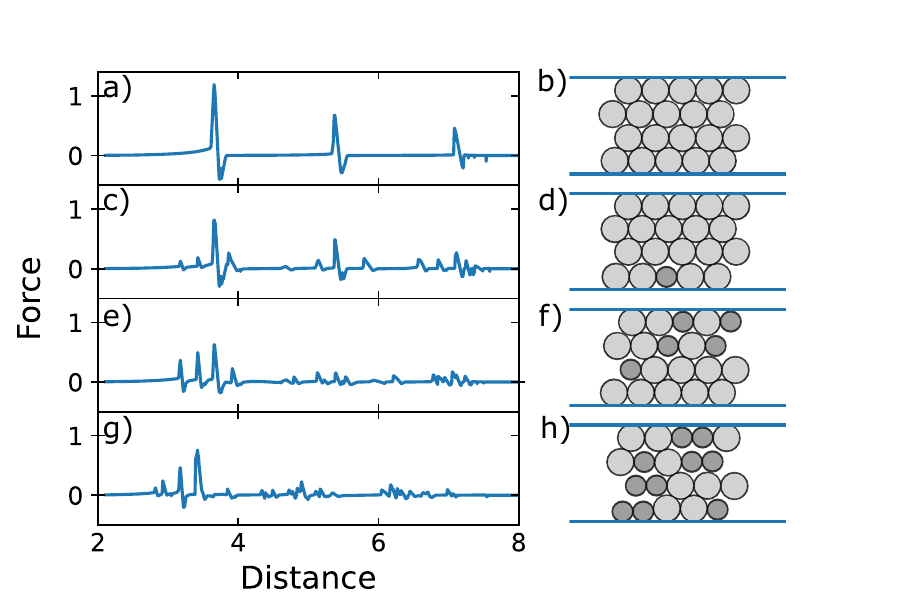}
\caption{\small {Evolution of the force profile as the defect fraction $\phi$ increases. The more bidisperse the aggregates becomes the noisier the force profile is; the individual fracture events involving many droplets split into a broad sequence of smaller fractures. The images in the right columns are snapshots from the state prior to compression of the system at distance 8.}}
\label{fig:force_bidispers}
\end{figure}

Figure \ref{fig:force_bidispers} shows how the force profile changes as we substitute differently sized particles in a 20 droplet array, becoming progressively more disordered as the fraction of defects rises from $\phi = 0$ to $\phi = 0.5$. (The compression process is shown in Movies S2 and S3 in the Supplemental Material \cite{Supplemental}).   
Figure \ref{fig:force_bidispers}(a) shows the force profile for the monodisperse droplet aggregate shown in Fig.~\ref{fig:force_bidispers}(b), which as discussed in the previous section shows clear force peaks connected to well-defined row reductions.  Introducing a single small droplet results in a force profile and droplet array shown in Figs.~\ref{fig:force_bidispers}(c,d).  This single defect causes the appearance of smaller peaks, signalling additional smaller fracture events and thus a more disordered row reduction.  Figures \ref{fig:force_bidispers}(e-h) show the force profile and initial droplet configuration for $\phi=0.25$ and $\phi=0.50$.  Introducing more defects introduces more small force peaks.  At a defect fraction of $\phi=0.5$ there are no distinct ``row reductions'', but rather a nearly continuous series of small fractures.

In the prior experimental work, Ono-dit-Biot {\it et al.} developed a predictive model for the number of peaks in the force profile for a compressed aggregate\cite{ono-dit-biot_Rearrangement_2020}: 
\begin{equation}
\frac{\Delta N(\phi)}{\Delta N_{\mathrm{peak}}}=2\sqrt{(1-\phi)\phi},
    \label{eq:maxpeaks}
\end{equation}
where $\Delta N(\phi)=N(\phi)-N(0)$ is the excess number of peaks $N(\phi)$ observed for a given defect concentration over the number of peaks $N(0)$ for the original aggregate ($N(0)$ is the number of starting rows minus one), and $\Delta N_{\mathrm{peak}}$ is a fitting parameter to the highest amount of peaks for a given droplet configuration.

\begin{figure}
\begin{center}
\includegraphics[scale=0.6, origin=c]{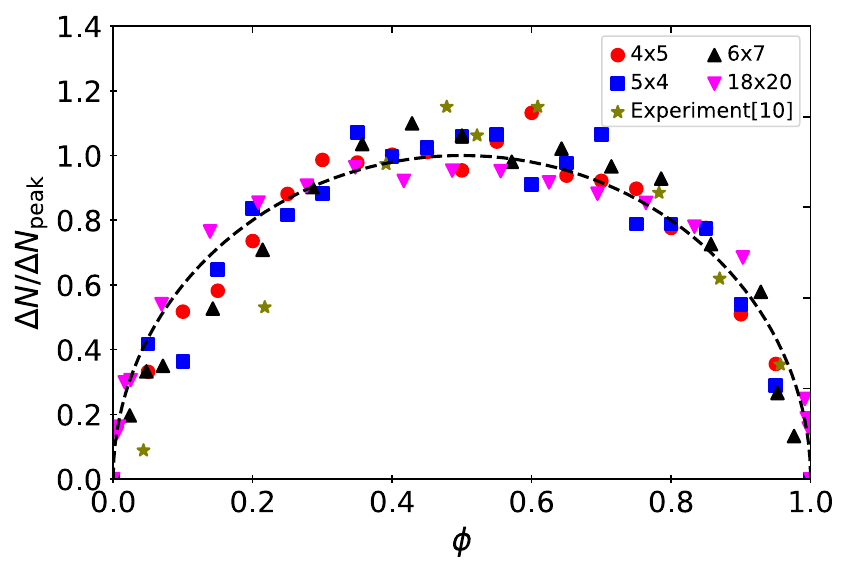}\caption{\small {$\Delta N(\phi)$ normalized by the fitting parameter $\Delta N_{\mathrm{peak}}$ for five different droplet arrays. The dashed line corresponds to the prediction of Eqn.~\ref{eq:maxpeaks}. Four of the data sets correspond to simulation data, while the points with the star marker are from the experimental data of Ono-dit-Biot {\it et al.} \cite{ono-dit-biot_Rearrangement_2020}. The experimental droplet array consists of $23$ particles in three columns containing 8, 7, and 8 particles.}}
\label{fig:normed_peaks}
\end{center}
\end{figure}

The prediction given by Eqn.~\ref{eq:maxpeaks} describes the simulation data well, as shown for four examples in Fig.~\ref{fig:normed_peaks} where the data have been scaled in each case by the $\Delta N_{\mathrm{peak}}$ that best fits each data set.  Furthermore, this data collapse agrees with the experimental results of Ref.~\cite{ono-dit-biot_Rearrangement_2020} (star symbols in Fig.~\ref{fig:normed_peaks}), and extends their $3 \times 8$ array results up to an $18\times 20$ droplet array.  Above this size, the peaks from individual fracture events begin to blur together around $\phi \approx 0.5$, making it challenging to correctly measure $N(\phi)$.

\begin{figure}
\begin{center}
\includegraphics[scale=0.6, origin=c]{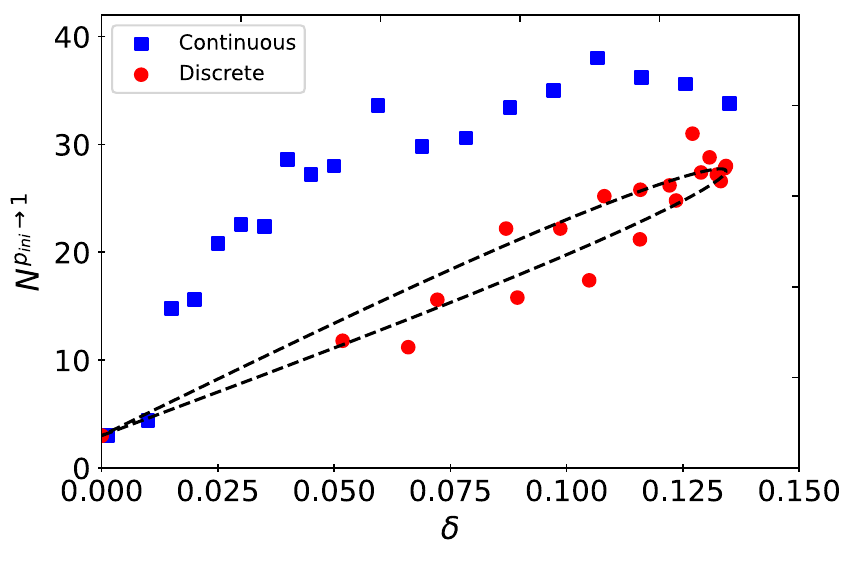}\caption{\small {$N^{p_{ini}\rightarrow 1}$ as a function of the polydispersity $\delta$ for the discrete $4\times 5$ bidisperse case used in Fig.\ref{fig:normed_peaks}, and a continuous size distribution. The amount of fractures events depends on the chosen size distribution, with the discrete case increasing somewhat lineally, while the continuous case grows nonlinearly.  The dashed line corresponds to the prediction of Eqn.~\ref{eq:maxpeaks}.  For the bidisperse case $\delta(\phi)$ is not symmetric between $\phi$ and $(1-\phi)$, so thus Eqn.~\ref{eq:maxpeaks} has two branches as shown \cite{meer2023}. }}
\label{fig:cont_disp}
\end{center}
\end{figure}

We can also consider how the bidisperse sample compares with a nominally single-component sample composed of polydisperse particles.  To do this, we use particles with sizes distributed according to a Gaussian, characterized by polydispersity $\delta$ defined as the standard deviation of the distribution divided by the mean.  Figure \ref{fig:cont_disp} shows the amount of fracture events occurring during the whole compression as a function of polydispersity.  The blue squares correspond to the Gaussian distribution, and the red circles correspond to the bidisperse distributions considered above, now plotted as a function of $\delta$ calculated from each distribution's standard deviation and mean size. For the discrete bidisperse data set, we have used the $4\times 5$ droplet case previously shown in Fig.~\ref{fig:normed_peaks}.  The continuous polydispersity case has many more fracture events than the bidisperse distributions with equivalent $\delta$.  Examining the individual movies, the increase in fracture events is because the case of continuous polydispersity acts to introduce weak points into the array in many locations simultaneously.  That is, the continuous polydispersity case is somewhat analogous to a bidisperse array with $\phi \approx 0.5$ and a size ratio that grows with increasing Gaussian width $\delta$.

\subsection{Competing Fractures}
\label{competing}

\begin{figure}
\includegraphics[scale=0.5, origin=c]{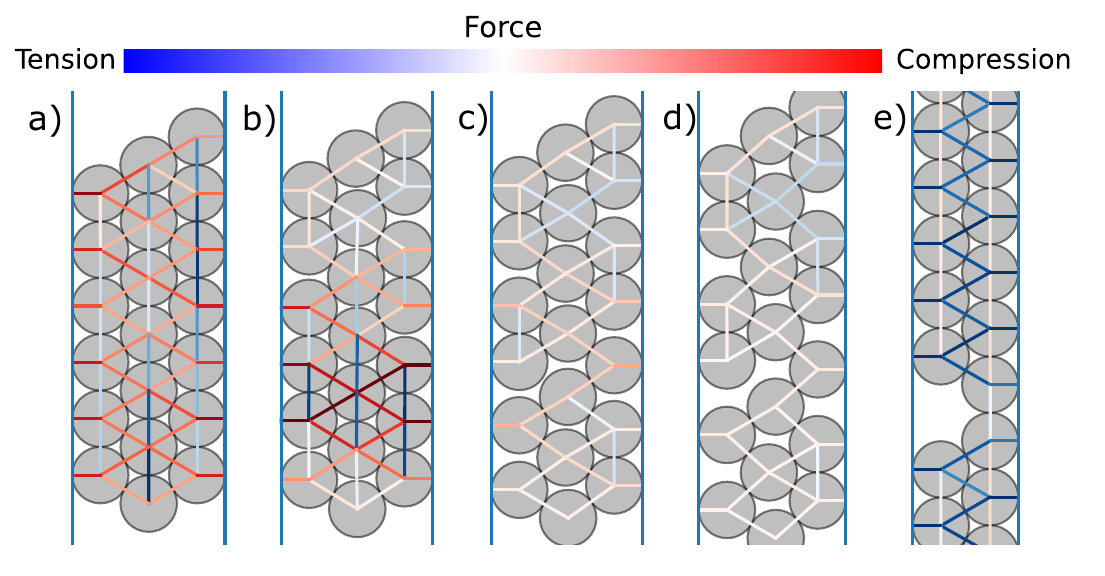}\caption{\small {  Successive images of of a 20 droplet aggregate undergoing a second row reduction, from the original 4 rows by 5 columns configuration.  By panel (c) we can see the formation of two fractures, which are misaligned as they propagate through out the array. Thus the collapse in (d) is disordered, and the final packing with a defect in (e).}}
\label{fig:comp_fracs}
\end{figure}

The larger disorder in the fracture process when adding defects is expected.  We additionally observe a new behavior in droplet aggregates even with low polydispersity not seen in Ref.~\cite{ono-dit-biot_Rearrangement_2020}:  competing fractures.  In Sec.~\ref{monodisperse} we focused on the compressed crystal undergoing single coordinated fractures resulting in a change from one hexagonal array to a smaller array with one fewer row.  However, sometimes two or more fractures nucleate at multiple sites in the array.  As the droplet raft is further compressed, these fractures propagate leading to misalignment:  the compressed array, upon completion of the fractures, is no longer hexagonal.  Instead, we see holes or other defects in the structure.  An example of competing fractures can be seen in Fig.~\ref{fig:comp_fracs}. For this particular example in the first snapshot [Fig.~\ref{fig:comp_fracs}(a)] the droplets are compressed throughout the whole array, with some variability due to the minimal underlying droplet polydispersity.  This pressure is alleviated by breaking depletion bonds, as seen in Fig.~\ref{fig:comp_fracs}(b):  but this occurs mainly on the upper portion of the array, while the lower half remains compressed.  Eventually the lower end starts another fracture event, but the second fracture does not align with the first fracture, as seen in Fig.~\ref{fig:comp_fracs}(c). This results in an disorderly row reduction as seen in the last snapshot Fig.~\ref{fig:comp_fracs}(d).

\begin{figure}
\includegraphics[scale=0.62, origin=c]{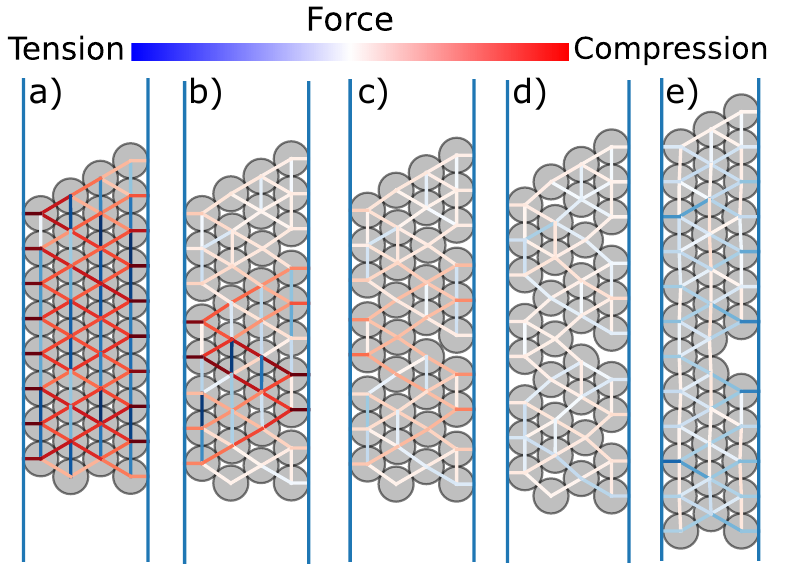}\caption{\small {Successive images of a 36 droplet aggregate, undergoing its third row reduction from the original six by six configuration.  By panel (c) we can see the formation of two fractures, which are misaligned as they propagate through out the array. Thus the collapse in (d) is disordered, and the final packing with a defect in (e).}}
\label{fig:comp_fracs36}
\end{figure}

A similar example is present in Fig.~\ref{fig:comp_fracs36} for a 36 droplet case. Upon compression, a fracture originates first at the ``top" of the array, with a secondary fracture nucleating later  at the ``bottom." Both fractures fail to align, causing again a disorderly collapse in the crystal.  Another example of competing fractures in large arrays is presented in Supplemental Material Movie S4, showing the compression of a 121 particle array \cite{Supplemental}.

\begin{figure}
\includegraphics[scale=0.5, origin=c]{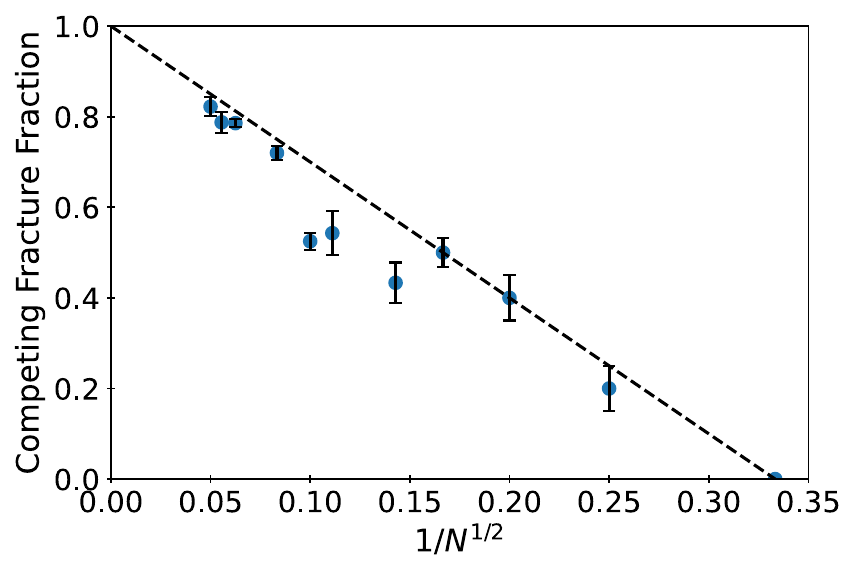}\caption{\small The fraction of row reductions observed to have competing fractures as a function of $1/N^{1/2}$, using the number of droplets $N$. The data correspond to initially square arrays such as the array in Fig.~\ref{fig:global_fracture}(a).  The error bars reflect the standard deviation over five runs.}
\label{fig:fracs_v_N}
\end{figure}

While Fig.~\ref{fig:comp_fracs} is a small droplet array, we would expect that in larger arrays there are more potential sites for fracture events to start.  Furthermore, even if a fracture starts in one location, it propagates to other locations at a finite speed:  it is possible that the fractures can't spread fast enough to cover the whole crystal before another fracture event is nucleated elsewhere.  To test this suggestion, we measure the fraction of row reductions that occur via competing fractures for different numbers of droplets $N$ and plot this in Fig.~\ref{fig:fracs_v_N}.  This fraction is defined by averaging over several simulations with the same $N$ but varying the random seed for the polydispersity.  In particular, for each simulation run we count the number of row reductions that have multiple competing fractures.  The one exception is that we ignore the very last row reduction ($2 \rightarrow 1$) which is always clean; thus if we start with $p$ rows, there are $p-2$ total row reductions which could potentially have competing fractures. Then we calculate the average fraction of row reductions with competing fractures over all runs with the same starting configuration. The data points are plotted as a function of $1/N^{1/2}$ and exhibit a fairly linear trend:  larger arrays have more competing fractures, with an extrapolation to all fracture events being competing fractures in the $N \rightarrow \infty$ limit.  The limit where no competing fractures occurs corresponds to a $3\times 3$ array ($N=9$) for which there is no longer a possibility for competing fractures; the array is too small to fit two fracture events.  Overall, Fig.~\ref{fig:fracs_v_N} confirms the basic idea, that larger arrays have more potential ways for competing fractures to occur.

\begin{figure}
\includegraphics[scale=0.5, origin=c]{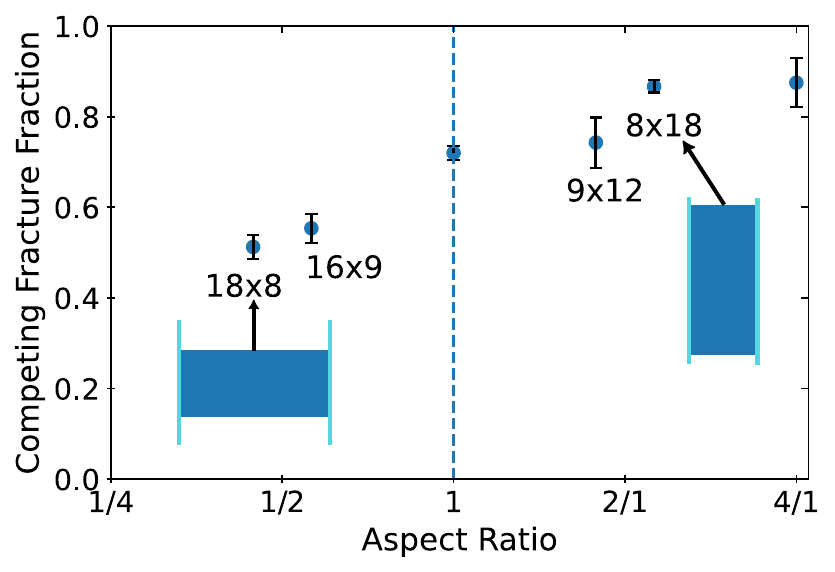}\caption{\small{ The fraction of competing fractures as a function of aspect ratio using $N=144$ droplet configurations. The aspect ratio is defined as $q/p$, with a low aspect ratio corresponding to ``tall'' configurations (many rows parallel to the walls, with the walls starting quite far apart), and high aspect ratios to ``wide'' configurations (few rows between walls that start close together).  The error bars reflect the standard deviation over five runs.}}
\label{fig:fracs_v_aspratio}
\end{figure}

We investigate how the presence of competing fractures is influenced by the initial droplet array aspect ratio, defined as $AR=q/p$.  The results are shown in Fig.~\ref{fig:fracs_v_aspratio} based on calculations with $N=144$ droplets.  A lower aspect ratio corresponds to a ``taller" initial configuration with many rows, and as expected the initial row reductions have little amount of competing fractures.  As the array becomes wider, competing fractures become more prevalent, similar to the wide array shown in Fig.~\ref{fig:comp_fracs}.  This confirms that for a wider configuration the compression from the wall at the far ends of the crystal can produce separate fracture events.  The data should be interpreted with caution:  the ``taller'' configurations with aspect ratio less than 1 will be compressed and pass through the ``wider'' configurations, and thus some number of the competing fractures observed for the taller configurations occur when the array is at a later compression stage and is thus wider.  This likewise is a factor in the data of Fig.~\ref{fig:fracs_v_N}, that the larger $N$ arrays sample higher aspect ratio configurations during their compression which have the higher propensity for competing fractures.

\begin{figure}
\includegraphics[scale=0.5, origin=c]{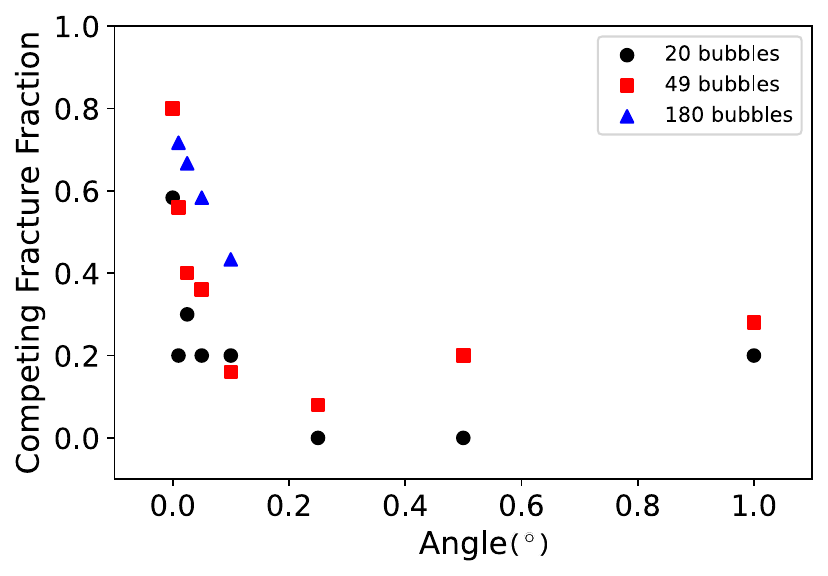}\caption{\small{ The fraction of competing fracture as a function of the angle between the two walls, for three different droplet array sizes as indicated.  The data suggest a slight angle between the walls can dramatically decrease the frequency of competing fractures.}}
\label{fig:fracs_v_angle}
\end{figure}

Competing fractures were not seen in the experiments of Refs.~\cite{ono-dit-biot_Rearrangement_2020,ono-dit-biot_Mechanical_2021}.  We speculate that this may be due to the experimental challenge of aligning the two glass pipettes that serve as the compressing walls.  To test this hypothesis, we study the dependence of competing fractures on the relative angle of the moving wall to the stationary wall, with $0^\circ$ representing perfectly parallel alignment.  The data are shown in Fig.~\ref{fig:fracs_v_angle} for arrays with 20 to 180 droplets as indicated.  As the angle of the wall increases, the frequency of competing fractures decreases, reaching a minimum close to an angle of $0.2^{\circ}$.  This is due to the wall compressing on one side of the array first, which results in fractures nucleating on that side first, and spreading throughout the crystal as the wall continues moving.  However, at still higher angles, close to the $0.2^{\circ}$ angle for the arrays shown in  Fig.~\ref{fig:fracs_v_angle},  the compression is sufficiently uneven that the number of rows between the walls becomes less well-defined, leading to an increase in competing fractures.  A tilt angle of $0.2^\circ$ is plausible for the experiments, and may have biased the experimental observations towards single fractures.

\section{Conclusion}

We have simulated a variety of two-dimensional arrays of droplets with attractive interactions as they undergo compression.  Inspired by the prior experimental work of Refs.~\cite{ono-dit-biot_Rearrangement_2020,ono-dit-biot_Mechanical_2021}, we reproduce and extend their key results.  First, we show how the effective elastic properties of the droplet arrays are related to the intrinsic spring constant acting between a pair of contacting droplets.  In particular, some of these springs are compressed and others are under tension; the tension bonds act to increase the overall spring constant exhibited by the droplet array.  Second, we confirm how the force required to initiate a fracture event scales with the aggregate size, extending the results to more droplets than the experiments studied.  Third, we find that intentionally adding defects into the otherwise hexagonally ordered array dramatically increases the number of fracture events, while at the same time decreasing the force required to initiate those fractures.  The dependence of the fracture events on defects is in quantitative agreement with the model developed in Ref.~\cite{ono-dit-biot_Rearrangement_2020}, and with the simulations we were able to extend the size of the cluster studied by an order of magnitude more droplets.  Of interest is that the number of excess fractures scales as $\sqrt{\phi}$ for a small fraction $\phi$ of defects.  The derivative of this diverges as $\phi \rightarrow 0$, indicating that for a perfect crystal, adding in any density of defects dramatically increases the ease of breaking the crystal.  The maximum disorder occurs when the sample is composed of an equal mixture of two sizes of droplets, in reassuring agreement with the decades of simulations which have used mixtures of equal numbers of two particle sizes to model glasses, for example in the classic papers of Kob and Andersen \cite{kob95a,kob95b}.

Our simulations also found a phenomenon not observed in experiments, which is the presence of competing fractures.  This occur when two independent fracture events start in different locations, and when they propagate through the sample, they do not match in the middle.  Competing fractures result in the post-fracture array being more disordered.  These are more prevalent for larger droplet arrays, giving some sense of why they might not have been observed in the experiments.  We demonstrated that if the two walls compressing the crystalline aggregate are slightly tilted with respect to each other, this helps bias the formation of cracks toward the more compressed side.  A tilt angle of $\sim 0.2^\circ$ is optimal in the simulations for suppressing competing fractures, and this is entirely plausible to have been present in the experimental work \cite{ono-dit-biot_Mechanical_2021}.  This also suggests that in real crystals undergoing compression, slight mis-alignment of compressing surfaces could affect how samples fracture.  Note that in our simulations, symmetry is broken by the slight polydispersity of droplet sizes -- introduced to match the experimental polydispersity.  Such polydispersity would not be present in ideal crystals.  Nonetheless, thermal fluctuations might facilitate multiple sites for fractures to be initiated.

\begin{acknowledgments}

This material is based upon work supported by the National Science Foundation under Grant Nos. CBET-1804186 and CBET-2002815 (P.E.I. and E.R.W.). 

\end{acknowledgments}


%

\end{document}